\documentclass[12pt,amsfonts]{article}

\usepackage{amsfonts}

\def\bn {\begin{eqnarray}}
\def\en {\end{eqnarray}}

\begin{document}


\title{
Cosmological perturbations
    in FRW model with scalar field within Hamilton-Jacobi formalism and symplectic projector method  }

\author{
    \textbf{Dumitru Baleanu}\thanks{E-mails:
dumitru@cankaya.edu.tr, baleanu@venus.nipne.ro} \\[12pt]
    Department of Mathematics and Computer Sciences, \\
    Faculty of Arts and Sciences, \\
    \c{C}ankaya University, 06530 Ankara, Turkey\\
    and\\
 Institute of Space Sciences, P.O. BOX, MG-23, R 76900,\\
Magurele-Bucharest, Romania }

\date{}   

\maketitle

\begin{abstract}
The Hamilton-Jacobi analysis is applied to the dynamics of the
scalar fluctuations about the Friedmann-Robertson-Walker (FRW).
The gauge conditions are found from the consistency conditions.
 The physical
degrees of freedom of the model are obtain by symplectic projector
method. The role of the linearly dependent Hamiltonians  and the
gauge variables in Hamilton-Jacobi formalism  is discussed.
\end{abstract}

\vspace*{2cm}


\newpage


\section{Introduction}

    The cosmological models which include the theory of a scalar field
coupled to gravity play an important role during the last period.
A transformation from a reparametrization-invariant system to an
ordinary gauge system was applied for deparametrizing cosmological
models.
  In the path integral approach to false
vacuum decay with the effect of gravity, there is an unsolved
problem, called the negative mode problem.
 A conjecture it was proposed that there should be no supercritical
supercurvature mode.This conjecture was verified for a wide
variety of tunnelling potentials \cite{takahiro}.
 For the monotonic potentials
no negative modes were reported about the Hawking-Turok instanton.
For a potential with a false vacuum the Hawking-Turok instanton it
was shown that we obtain a negative mode for certain initial data
\cite{HT}.
 It was shown that the cosmological perturbations in
Lorentzian regime are related to the cosmic microwave background
radiation and large scale structure formation
\cite{mukhanov,garriga98,gratton}.
 The unconstrained reduced action corresponding to the dynamics
of scalar fluctuations about the FRW background was obtained by
applying  Dirac's method of singular Lagrangian systems
\cite{dirac,henneaux}.
  The results were applied to the negative
mode problem in the description of tunnelling transitions with
gravity \cite{khved}. There are several known methods in obtaining
and dealing with unconstrained quadratic action in terms of the
physical variables \cite{garriga98,gratton,laruti} in the theory
of scalar field coupled to gravity in non-spatially flat FRW
Universe but the main problem appears at the quantum level
\cite{khved}.
 For these reasons new quantization methods as Hamilton-Jacobi
 method
 and the symplectic projector method \cite{projector1,projector2,projector3,projector4}
  should be applied
on the theory mentioned above.
  By adding a surface term to the action functional
the gauge invariance of the systems whose
Hamilton-Jacobi equation is separable was improved
\cite{simeone2}.

  Hamilton-Jacobi formalism
(HJ) based on Carath\'{e}odory's idea \cite{caratheodory} gained a
considerable importance during the last decade due to its various
applications to quantization of constrained systems
\cite{pimentel98}.

  However, some difficulties may occur
for HJ in dealing with linear dependent constraints. The main
problem comes from the construction of the canonical Hamiltonian.
Let us assume that the canonical Hamiltonian is a linear
combination of two terms and the second one is proportional to a
given field having its momentum zero. After imposing the
integrability condition for that momentum we obtain a new
constraint, therefore the canonical Hamiltonian is a linear
combination of two constraints.\\
 Therefore one of interesting and
yet not solved question is how to deal with the total differential
equations within HJ in the above mentioned case.

Another issue is related to the gauge fixing procedure within  HJ
formalism. Can we find inside of HJ a mechanism to obtain the
gauge fixing condition?

In order to analyze the above mentioned open problems we have to
apply HJ formalism to a constrained system possessing linearly
dependent constraints.

{}For these reasons the application of HJ formalism to
cosmological perturbations in FRW model with scalar field is an
interesting issue.

The paper is organized as follows:

In Section 2 the model is presented. Section 3 presents briefly HJ
formalism. The symplectic projector method is discussed in Section
4. In Section 5 the gauge fixing conditions of the investigated
model are discussed inside HJ formalism and the true degrees of
freedom are obtained within symplectic projector method. Finally,
Section 6 is dedicate to our conclusions.

\section{ The model}
The action of the system of scalar matter field coupled to gravity
is given by
\begin{equation}
S=\int d^4x\sqrt{-g}
        \left[\frac{R}{2k}
        -\frac{1}{2}\nabla_{\mu}\phi\nabla^{\mu}\phi-V(\phi)
        \right].
\end{equation}
  Here $k=8\pi G$ represents the reduced Newton's constant and the
scalar potential field is denoted by $V(\phi)$ \cite{khved}.

By expanding the metric and the scalar field over an FRW type
background one obtains
\begin{eqnarray}
ds^2 &=& a(\eta)^2 \left[ - ( 1 + 2A(\eta) Y)d\eta^2
     + 2\mathcal{B}(\eta)Y_{\mid i}d\eta dx^i
        \right.
   \nonumber \\[1mm]
     & & \left.  \quad  \quad \quad
          + \{ \gamma_{ij} ( 1 - 2\Psi(\eta) Y)
          + 2\mathcal{E}(\eta)Y_{\mid ij}dx^idx^j \} \right],
  \nonumber \\[2mm]
\phi &=& \varphi(\eta) + \Phi(\eta)Y.
\end{eqnarray}
  Here $\gamma_{ij}$ represents the three-dimensional metric on the
constant curvature space sections, $a$ and $\phi$ denote the
background field values and $A,\Psi,\Phi,{\cal B}$ and ${\cal E}$
are small perturbations. In addition, Y represents a normalized
function of 3-dimensional Laplacian, $\Delta Y=-k^2Y$, and
vertical line denotes the covariant derivative with respect to
$\gamma_{ij}$.

The background fields a and $\phi$ are subjected to the following
equations
\begin{eqnarray}\label{ee}
{\cal H}^2-{\cal H}^{'}+{\cal K}
&=&  \frac{k}{2}{\phi^{'}}^2,\cr
2{\cal H}^{'} +{\cal H}^2 +{\cal K}
  &=&
\frac{k}{2}\left(-{\phi^{'}}^2+2a^2V(\phi) \right)  \cr
\phi^{''}+2{\cal
H}{\phi^{'}}+ a^2\frac{\delta V}{\delta\phi}
&=& 0.
\end{eqnarray}
In (\ref{ee}) the prime denotes a derivative with respect to
conformal time $\eta$, ${\cal H}=\frac{a^{'}}{a}$ and ${\cal K}$
denotes the curvature parameter which takes the values 1,0,-1 for
closed, flat and open universes, respectively.

The total action containing only the second order terms becomes
\begin{equation}
S=S^{(0)}+S^{(2)},
\end{equation}
where $S^{(0)}$ represents the action of the background solution
and $S^{(2)}$ is  quadratic in perturbations. The corresponding
Lagrangian is as follows
\begin{eqnarray}\label{lagran}
{\cal L} \!\!  &=&   \!\!   
   \frac{a^2\sqrt{\gamma}}{2k}
    \biggl[ -6\Psi^{'2} + 2(k^2 - 3{\cal K})\Psi^2
        + k\biggl\{\Phi^{'2}
             - (a^2\frac{\delta^2V}{\delta\phi\delta\phi}
        + k^2)\Phi^2 + 6\phi^{'}\Psi^{'}\Phi \biggr\}  \cr
   && -
    \biggl\{
                2k{\phi^{'}\Phi^{'}} +2ka^2\frac{\delta V}{\delta\phi}\Phi
        +12{\cal H}{\Phi}^{'}+4(k^2-3{\cal K})\Psi
        \biggr\}A \cr
 &&  -  2 ( {\cal H}^{'}+2{\cal H}^2-{\cal K}) A^2 \biggr].
\end{eqnarray}
\section{ Hamilton-Jacobi formalism}
HJ formalism presented in this section is based on
 Carath\'{e}odory's idea of equivalent Lagrangians
\cite{caratheodory}. This approach can be considered an alternative
method of quantization of constrained systems and it was subjected
under an intense debate during the last decade
\cite{pimentel98,samulih,baleanujma,baleanumpla,baleanu2001,pimentel2004,pimentel2003,
baleanu2004}. The starting point of this method is a singular
Lagrangian $\mathcal{L}$, therefore the corresponding  Hessian
matrix is singular. In this case a set of primary constraints
appears naturally. Instead of working with one Hamiltonian, in this
method we use the initial canonical Hamiltonian $H_0$ and all
primary constraints $H_\alpha^{'}$.
  Namely, the "Hamiltonians" are
\begin{equation}\label{ham1}
H_\alpha^{'}=p_\alpha +H_{\alpha}(t_\beta,q_a,p_a),
\end{equation}
where $\alpha,\beta=n-r+1,\cdots,n a=1,\cdots,n-r$
and the canonical one
\begin{equation}\label{ham2}
H_0 = p_a w_a +{\dot q_\mu}p_{\mu} \mid_{p_\nu=H_\nu}
    -L(t,q_i,{{\dot  q}_{\nu},{{\dot q}_a}=w_a),
    \, \, \  \nu=0,n-r+1,\cdots,n.}
\end{equation}
Using (\ref{ham1}) and (\ref{ham2}) a set of total differential
equations is obtained
\begin{equation}\label{sys}
 dq_a = \frac{\partial H_\alpha^{'}}{\partial p_a}dt_\alpha,  \
dp_a = -\frac{\partial H_\alpha^{'}}{\partial q_a}dt_\alpha,  \
dp_\mu=-\frac{\partial H_\alpha^{'}}{\partial t_\mu}dt_\alpha,
    \quad   \mu=1,\cdots,r,
\end{equation}
together with HJ function $z$ is defined by
\begin{equation}
dz=\left(
    -H_\alpha
    + p_\alpha\frac{\partial H_\alpha^{'}}{\partial p_\alpha}
    \right) dt_\alpha,
\end{equation}
where $t_\alpha$ are gauge variables \cite{caratheodory}.
The next step is to investigate the integrability of the system
(\ref{sys}).

On the surface of constraints the system of differential equations
is integrable  if and only if
\begin{equation}\label{integr}
[ H_\alpha^{'},H_{\beta}^{'}]=0.
\end{equation}
The difficulties appear in this formalism when the Hamiltonians
are not in the form (\ref{ham1}). In this case the physical
significance from HJ point of view is lost.Therefore, we have to
make a canonical transformation to be able to recover the physical
significance. Finding a suitable canonical transformation for a
given constrained system is not an easy task in general. The
surface terms play an important role in finding the integrability
conditions in HJ formalism.

 If (\ref{integr}) is not fulfilled,then another set of "Hamiltonians"
 arises and we  subject
them to the integrability conditions. The process ends when no new
"Hamiltonian" appears.

\section{Symplectic Projector Method}

Let us assume that a system admits only second constraints

\begin{equation}
\phi^m(\zeta^M)=0,
\end{equation}
where $\zeta^M=(x^a,p^a)$, $M=1,2,\cdots...2N$ are the
coordinates.

The definition of symplectic projector is the following (for more
details see \cite{projector1,projector2,projector3, projector4}
and the references therein )

\begin{equation}\label{proje}
\Lambda^{MN}=\delta^{MN}-J^{ML}\frac{\delta\phi_m}{\delta\zeta^L}\Delta_{mn}^{-1}\frac{\delta\phi_n}{\delta\zeta^N},
\end{equation}
where $\Delta_{mn}^{-1}$ represents the inverse of the following
matrix

\begin{equation}
\Delta_{mn}=\{\phi_m,\phi_n\}
\end{equation}
and $J^{MN}$ denotes the symplectic two form. Thus, the action of
the symplectic projector defined by (\ref{proje}) is to project
$\zeta^{M}$ onto local variables on the constraint surface defined
below
\begin{equation}
\zeta^{{\star} M}=\Lambda^{MN}\zeta^N.
\end{equation}

\section{ The model and the gauge fixing conditions  \\
        within HJ formalism }
{}From the Lagrangian density (\ref{lagran}) we
obtain the canonical momenta as
\begin{eqnarray}
\Pi_{\Psi} &=& \frac{6a^2\sqrt{\gamma}}{k}(-\Psi^{'}
        +\frac{k}{2}\psi^{'}\Phi-{\cal H}A)
    \nonumber \\
\Pi_{\Phi}
         &=&  a^2\sqrt{\gamma}(\Phi^{'}-{\phi}^{'}A),
    \label{mnm} \\
\Pi_A  &=& 0.
    \label{has1}
\end{eqnarray}
{}From (\ref{has1}) we conclude that A is a gauge variable and
that $H_1^{'}=\Pi_A$ represents a "Hamiltonian". By using
(\ref{mnm}) and (\ref{has1}) the canonical Hamiltonian becomes

\begin{eqnarray}\label{canonicalH}
H_C &=& -\frac{k}{12a^2\sqrt{\gamma}}\Pi_{\Psi}^2
    +\frac{1}{2a^2\sqrt{\gamma}}\Pi_{\Phi}^2
    +\frac{k}{2}\phi^{'}\Pi_{\Psi}\Phi\cr
  &+&
    a^2\sqrt{\gamma}[-\frac{k^2-3{\cal K}}{k}\Psi^2
    +\frac{1}{2}(a^2\frac{\delta^2 V}{\delta\phi\delta\phi}
    -\frac{3}{2}k{\phi}^{'2}+k^2)\Phi^2]\cr
 &+&
       A\{{\phi^{'}}\Pi_{\Phi}-{\cal H}\Pi_{\Psi} +a^2\sqrt{\gamma}
    [(a^2\frac{\delta V}{\delta\phi}+3{\phi}^{'}{\cal H})\Phi
    +\frac{2(k^2-3{\cal K})}{k}\Psi]\},
\end{eqnarray}
therefore
\begin{equation}\label{hass}
H_0^{'}=p_0+H_C.
\end{equation}
The next step in HJ formalism is to obtain the total differential
equations by using (\ref{has1}) and (\ref{hass}). In our case we
obtain the following set of total differential equations
\begin{eqnarray}
d\Psi
   &=&-\frac{k\Pi_{\Psi}d\tau}{6a^2\sqrt{\gamma}}
    +\frac{k}{2}{\phi^{'}}\Phi d\tau - A{\cal H}d\tau,
 \label{p1}
 \\
  d\Phi
     &=& (\frac{\Pi_{\phi}}{a^2\sqrt{\gamma}}+A\phi^{'})d\tau,
 \label{p2}
\\
d\Pi_{\Psi}
    &=&  2 a^2\sqrt{\gamma}\frac{k^2-3{\cal K}}{k}\Psi
    d\tau-\frac{2a^2\sqrt{\gamma}A(k^2-3{\cal K})}{k}d\tau,
\label{p3}
\\
d\Pi_{\Phi}
 &=&  -\frac{\Pi_{\Phi}}{a^2\sqrt{\gamma}}d\tau-
      A a^2\sqrt{\gamma}(\frac{a^2\delta V}{\delta\phi}
    + 3{\phi}^{'}{\cal H})d\tau
 \nonumber \\
& & - a^2(\frac{\delta^2 V}{\delta\phi\delta\phi}
    - \frac{3}{2}k{\phi}^{'2}+k^2)\Phi d\tau. \label{p4}
\end{eqnarray}
Taking into account (\ref{canonicalH}) and the consistency
condition
\begin{equation}\label{unu}
d\Pi_A=0,
\end{equation}
we get  a new "Hamiltonian" denoted by $H_2$. Namely, the form of
$H_2$ is given by
\begin{equation}\label{doi}
H_2= {\phi^{'}}\Pi_{\Phi}-{\cal H}\Pi_{\Psi}
+a^2\sqrt{\gamma}[(a^2\frac{\delta V}{\delta\phi}+3{\phi}^{'}{\cal
H})\Phi +\frac{2(k^2-3{\cal K})}{k}\Psi].
\end{equation}
In order to close the chain the variation of $H_2$ must be zero,
otherwise a new constraint will appear. By using (\ref{p1}),
(\ref{p2}), (\ref{p3}) and (\ref{p4})we may find after some
tedious calculations that if
\begin{equation}\label{gauge}
A-\Psi=0, \quad  \Pi_{\Psi}=0,
\end{equation}
then $dH_2=0$ provided that A is given as a function of background
fields,$\Phi$ and $\Pi_{\Phi}$.

\subsection{Physical Hamiltonian}

To find the true degrees of freedom of the proposed model we used
the symplectic projector method
\cite{projector1,projector2,projector3,projector4} . The set of
second class constraints to start with is as follows

\begin{eqnarray}\label{coni}
&C_1&={\phi^{'}}\Pi_{\Phi} +a^2\sqrt{\gamma}
    [(a^2\frac{\delta V}{\delta\phi}+3{\phi}^{'}{\cal H})\Phi
    +\frac{2(k^2-3{\cal K})}{k}\Psi]\cr
    &C_2&=\Pi_A,C_3=\Pi_{\psi},C_4=A-\Psi.
\end{eqnarray}

By using ({\ref{coni}}) we obtain the form of matrix $\Delta$ as
follows

\begin{equation}\Delta=\left(
\begin{array}{cccc}
0&0&2\frac{a^2\sqrt{\gamma}(k^2-3K)}{k}&0\\
0&0&0&-1\\
-2\frac{a^2\sqrt{\gamma}(k^2-3K)}{k}&0&0&1\\
0&1&-1&0
\end{array}
\right)\delta({\vec{x}-\vec{y}})
\end{equation}
The form of the matrix projector becomes
 \begin{equation}\Lambda = \left(
\begin{array}{cccccc}
0 &-k\frac{a^2\frac{\delta V}{\delta\phi}+3{\phi^{'}}{\cal H}}{2(k^2-3K)} &0&0&-\frac{k\phi^{'}}{2a^2\sqrt{\gamma}{(k^2-3K)}}&0\\
0 &1 &0&\frac{k\phi^{'}}{2a^2\sqrt{\gamma}{(k^2-3K)}}&0&\frac{k\phi^{'}}{2a^2\sqrt{\gamma}(k^2-3K)}\\
0 &-k\frac{a^2\frac{\delta V}{\delta\phi}+3{\phi^{'}}{\cal H}}{{2(k^2-3K)}} &0&0&-\frac{k\phi^{'}}{2a^2\sqrt{\gamma}{(k^2-3K)}}&0\\
0 &0 &0&0&0&0\\
0 &0 &0&-k\frac{a^2\frac{\delta V}{\delta\phi}+3{\phi^{'}}{\cal H}}{2(k^2-3K)}&1&-k\frac{a^2\frac{\delta V}{\delta\phi}+3{\phi^{'}}{\cal H}}{{2(k^2-3K)}}\\
0 &0 &0&0&0&0
\end{array}
\right)\delta({\vec{x}-\vec{y}})\end{equation} We observed that
$Tr \Lambda=2$, therefore we have only two true physical degrees
of freedom. Let us introduce the phase space vector $\xi$ with the
following components
\begin{equation}
(\xi^1,\xi^2,\xi^3,\xi^4,\xi^5,\xi^6)=(A,\Phi,\Psi,\Pi_A,\Pi_{\Phi},\Pi_{\Psi})
\end{equation}

By using (14) we obtain
\begin{eqnarray}
\zeta_1^{\star}&=&-k\frac{(a^2\frac{\delta
V}{\delta\phi}+3{\phi^{'}}{\cal
H})}{2(k^2-3K)}\xi^2-\frac{k\phi^{'}}{2a^2\sqrt{\gamma}{(k^2-3K)}}\xi^5,\cr
\zeta_2^{\star}&=&\xi^2
+\frac{k\phi^{'}}{2a^2\sqrt{\gamma}{(k^2-3K)}}(\xi^4+\xi^6)\cr
\zeta_3^{\star}&=&-k\frac{(a^2\frac{\delta
V}{\delta\phi}+3{\phi^{'}}{\cal
H})}{2(k^2-3K)}\xi^2-\frac{k\phi^{'}}{2a^2\sqrt{\gamma}{(k^2-3K)}}\xi^5,\cr
\zeta_4^{\star}&=&0,\cr
 \zeta_5^{\star}&=&-k\frac{(a^2\frac{\delta
V}{\delta\phi}+3{\phi^{'}}{\cal H})}{2(k^2-3K)}(\xi^4+\xi^6)
+\xi^5,\cr \zeta_6^{\star}&=0&.
\end{eqnarray}

We observed that

\begin{equation}\label{unu}
\xi_1^{\star}=\xi_3^{\star}
\end{equation}
and

\begin{equation}\label{doi}
\zeta_5^{\star}=-\frac{a^2\sqrt{\gamma}}{\varphi^{'}}\{\zeta_2^{\star}(a^2\frac{\delta
V}{\delta\phi}+3{\phi^{'}}{\cal
H})+\frac{1}{2(k^2-3K)}\zeta_1^{\star}\},
\end{equation}
therefore only two physical variables
$\zeta_1^{\star},\zeta_5^{\star}$ can be used as a starting point
for the quantization of the system. As it can be seen from
(\ref{unu}) and (\ref{doi}) we obtain the same degrees of freedom
as in \cite{khved}.

\section{Conclusions}

The integrability of HJ total differential equations is an open
and  attractive issue. In our study we obtained the gauge
conditions directly from the consistency conditions within HJ
formalism. This result is based on the fact that if the canonical
Hamiltonian represents a sum of two terms, the second one becomes
another "Hamiltonian" in HJ formalism. In other words the
canonical Hamiltonian represents a case of an irregular
Hamiltonian. If we denote $H_3=A-\Psi$ and $H_4=\Pi_{\Psi}$ we
obtain four "Hamiltonians" in our case. As it can be seen, the
obtained "Hamiltonians" are not in involution, therefore the
systems corresponding to these "Hamiltonians" is not integrable.
To make it integrable  we work on the surface of constraints and
this way leads us to the same canonical Hamiltonian from  up to a
constant.

The above result can be generalized for the case when the
canonical Hamiltonian has the form $H_c=H_0 +\phi_1H_1 +\cdots
\phi_nH_n$, where the fields $\phi_1,\cdots \phi_n$ do not appear
in any "Hamiltonians" $H_1,\cdots H_n$. In this case all $\phi_1,
\cdots \phi_n$ are gauge variables and they can be fixed after
imposing the integrability conditions. In order to calculate the
action we have to find the linearly independent "Hamiltonians"
possessing the physical significance from HJ point of view.

Since the set of four "Hamiltonians" is a second class typed in
Dirac classifications the symplectic projector method was used to
obtain the true degrees of freedom of investigated model. The
results were found to be in agreement with those from
\cite{khved}.

\section*{Acknowledgments}

 This work is partially supported by the Scientific and
Technical Research Council of Turkey.


\end{document}